# Accessing Phase Slip Events in Nb Meander Wires

*Deepika Sawle[1,2], Sudhir Husale[1,3], Sachin Yadav[1,2], Bikash Gajar[1,2], V. P. S. Awana[1,2] and Sangeeta Sahoo[1,2]\**

[1]*Academy of Scientific and Innovative Research (AcSIR), AcSIR Headquarters CSIR-HRDC Campus, Ghaziabad, Uttar Pradesh, 201002, India.*

[2]*Electrical & Electronics Metrology Division, National Physical Laboratory, Council of Scientific and Industrial Research, Dr. K. S Krishnan Road, New Delhi-110012, India.*

[3]*Indian Standard Time Metrology Division, National Physical Laboratory, Council of Scientific and Industrial Research, Dr. K. S Krishnan Road, New Delhi-110012, India.*

*\*Correspondences should be addressed to S. S. (Email: sahoos@nplindia.org)*

**Abstract:**

We report transport studies through Nb-based superconducting meander wires fabricated by focused ion beam (FIB) milling technique. The effect of meandering on quantum transport has been probed experimentally by a direct comparison with the pristine thin-film device before meandering. The normal metal (NM) to superconductor (SC) phase transition becomes a wide and multi-step transition by meandering. Below the transition temperature ($T_c$), the resistance-versus-temperature measurements reveal resistive tailing which is explained by the thermally activated phase slip (TAPS) mechanism. The TAPS fit indicates a selective region of the meander to be responsible for the resistive tailing. Besides, the phase slip (PS) mechanism in the meander is evident in its current-voltage characteristics that feature the stair-case type intermediate resistive steps during the SC-NM transition. The modulation of the intermediate resistive steps is investigated with respect to temperature and external magnetic field. It is observed that the PS events are facilitated by magnetic fields up to about 250 mT. Further, the critical current varies strongly on the temperature and magnetic field for $T < 0.5T_c$ and $H \leq 100$ mT where it fluctuates in an oscillatory manner. Finally, Nb based meander structures can be promising candidates for future PS based studies and applications.

**1. Introduction:**

Quantum fluctuations in 1D superconductor manifest as quantum phase slip (QPS) that leads to dissipation in superconducting nanowires [1]. In reduced dimension, phase of the superconducting order parameter fluctuates continuously and a phase change of '2π' in 1D nanowire causes momentary blockage in the flow of superconducting current carriers at a particular region of the nanowire and the region becomes resistive [2]. This phenomenon is called phase slip (PS) and the region where the PS occurs is called a phase slip centre (PSC) which is of the order of the coherence length of the superconductor. QPS and TAPS are the two main PS mechanisms that are controlled by quantum mechanical tunnelling and thermal fluctuations, respectively [3]. At temperatures far below the superconducting critical temperature ($T_c$), QPS is the main dissipation mechanism as the thermal fluctuations can be ignored, whereas, TAPS takes over at the vicinity of the $T_c$ where the effects of thermal fluctuations are pronounced. Further, QPS is known as the dual to the Josephson Effect (JE) and it is predicted that under appropriate electromagnetic environment with microwave excitation, quantized current steps are expected from a QPS junction leading to the much-awaited quantum current standard in the similar fashion as that occurred for quantum voltage standard from the JE [4].

In 2D superconductors, dissipation occurs mainly through superconducting vortices and phase fluctuation of the order parameter. With the application of longitudinal current, due to self-induced magnetic field, vortex-antivortex (V-Av) pairs are nucleated at the edges and/or in any weak points of the samples and start to move in the direction perpendicular to the current direction [5]. As current increases, under the influence of Lorentz force, the movement of the vortices gets faster and the slow-moving vortices transform into fast-moving vortices known as PSL state. Correspondingly, a sudden jump in voltage appears in the IVCs leading to the intermediate resistive states before the complete transition to the metallic state. These jumps actually relate to the PSLs and are formed across the width of the superconductor [6]. Commonly, wide transition and resistive tailing in temperature dependent resistance R(T) measurements indicate the presence of possible PS effects in superconducting nanostructures [7]. Further, the appearance of intermediate resistive steps in current voltage characteristics (IVCs) is the fingerprint of the PSLs in 2D [8-10]. Therefore, in wide superconducting strips, the resistivity appears due to vortex motion and phase slip events and the dynamics of V-Av pairs are mainly controlled by the bias current in absence of any external magnetic field [11, 12]. Further, the PSLs can be controlled and monitored with the external parameters like temperature and external magnetic field. For instance, under the influence of an external magnetic field in type-II superconductors, superconductivity can sustain through the quantized magnetic vortices that eventually lead to dissipative state when they are driven by an electrical current [13].

Disordered superconductors in dirty limit with high normal state resistance ($R_N$) are the promising candidates for PS studies [14]. Recently, it has been shown that the transition metal nitrides are good candidates for PS studies [15, 16]. Further, dilute magnetic doping is shown to trigger the PS events in NbGd films [8, 17]. However, the PS rate strongly depends on the dimension and the normal state resistivity [18]. For example, long uniform 1D nanowire is ideal for QPS junctions as reduced dimension leads to increased disorder and hence the increased $R_N$ [19-21].

One of most common and widely explored nano-structures is the meander structure which eventually can accommodate a reasonably long nanowire in a very precise area. Reports related to QPS and TAPS on meander structures include NbN meanders [22-24], FIB fabricated W-meanders [25], NbTiN [26] etc. There are only a very few reports available on PS effects in Nb meander structure and they mainly discussed the resistive tailing through *R(T)* measurements [27]. Here, we report of Nb meander structures fabricated by FIB milling to study PS mechanism through detailed transport measurements with respect to temperature and magnetic field. By FIB milling, on-chip fabrication of meander structures from a selected part of any existing thin-film device can be performed so that the comparison in the transport properties with respect to the dimensional change can be probed directly. Previously, we have reported that Nb thin-films with thickness 8-50 nm do not show the PS related features that generally appear in the *R(T)* and IVCs [8, 15]. However, wide transition along with resistive tailing in R(T) and staircase type resistive steps in IVCs imply the dominance of PS effects in the meander structures. Further, the *R(T)* is explained by the TAPS mechanism which indicates that a selective region of the meander wire might act as the core centre of the phase slip events that contribute mostly to the resistive tailing. In addition, the effects of temperature and applied magnetic field on IVCs have been investigated in detail and they act in a complimentary manner. Finally, the variation of critical currents with temperature and magnetic field has been investigated and the critical current is observed to depend strongly on temperature and magnetic field, particularly, at their low values where the critical current shows oscillatory type non-trivial fluctuations.

## 2. Experimental methods:

The Nb thin films were grown in an ultra-high vacuum (UHV) dc magnetron sputtering system by using a high purity (99.99%) Nb sputtering target of diameter 1.5". The films were deposited directly on pre-cleaned Si (100) substrate covered with thermally oxidized $SiO_2$ layer of 300 nm thickness which acts as the dielectric spacer between the film and the substrate. The sputtering was performed at room temperature with a base pressure less than $5 \times 10^{-9}$ Torr and in an Ar (99.9999% purity) environment at about 3 mBar. The deposition rate was ~ 3 Å/sec and the thickness of the Nb films reported in this study was about $(50 \pm 5)$ nm.

The Nb thin-films based multiterminal devices were fabricated using shadow mask. Initially, the electrical contact pads of Au (100 nm)/Ti (20 nm) layers were defined on the substrate. Next, the connecting superconducting Nb channel of length about 1300 micron and width of about 100 micron was sputtered deposited by using a complimentary mask. In order to avoid any oxidation of Nb films while exposing to air, finally a capping layer of 10 nm Si was deposited *in-situ* at the top of the Nb film. At the center of the device [Fig. 1 and Fig. S1 in the SI], Nb thin film was patterned into meander structure over a large area of about 230 $\mu m^2$ by FIB milling using a dual beam FIB microscope (Auriga Zeiss) equipped with field emission scanning electron microscope.

A Physical Properties Measurement System (PPMS) equipped with 14 T magnet by Quantum Design was used to perform the low temperature transport measurements. The devices were mounted on specific pucks and soldered for the electrical measurements. On a same device, transport measurements were carried out before and after the meandering. The *R(T)* measurements and IVCs were carried out using conventional four-probe geometry in current biasing mode. For *R(T)* measurements, low frequency (17 Hz) ac excitation of 500 nA (10 μA) was used for meander structure (initial thin film samples). Here, note that while transferring the sample among the instruments, the sample was exposed to air and hence the probable oxidation at the sidewall of the meander cannot be overruled. Further, an outer coating of Ga on the side wall can be expected for the meander wire and hence the effective width of the Nb meander gets reduced[28, 29]. Note that the meander structure was never exposed to Ga imaging and Ga ions were scanned only on the milled area[25].

## 3. Results and Discussions:

Fig. 1(a) displays the scanning electron microscopy (SEM) image of a device in which Nb thin-film serves as the superconducting channel connected with Au/Ti based electrical contacts. After initial transport characterization, the center part of the Nb thin-film was patterned into the meander structure which is shown in Fig. 1(b). The width and the gap between the lines in the meander is about 200 nm each as shown in the magnified image of a selected segment at the center of Fig. 1(b). The equivalent circuit, mainly consisting of resistors connected in series, is shown in Fig. 1(c). In the device geometry, the meander at the center is connected to the leads through Nb thin-film. Accordingly, the resistor $R_2$ in Fig. 1 (c) represents the meander wire, whereas $R_1$ relates to the resistance of the thin-film. As the width of the meander lines (200 nm) is much smaller than that of the thin film (100 μm), $R_2$ is expected to be much higher than $R_1$. Hence, the total resistance of the circuit is dominated by $R_2$ and the electrical transport is mainly governed by the meander structure.

The $R(T)$ measurements were carried out in four-terminal geometry down to 2K. Fig. 2(a) presents the comparison in $R(T)$ characteristics measured before and after the meander fabrication and are named as the film and the meander device, respectively. First, a sharp NM-to-SC transition occurs for the film whereas, the meander offers a broad and multistep transition.

Second, the $R_N$ for the meander is about three orders of magnitude more than that of the film. The inset of Fig. 2(a) represents the comparison of normalized resistance for the film and the meander where the normalization is done with respect to the resistance at 300 K. The residual resistance ratio (RRR), defined as the ratio of the resistance measured at 300 K and at the temperature (here 10 K) just above the onset of the metal-superconductor transition, is more for the film ($RRR_{film} \sim 3$) than that of the meander ($RRR_{meander} \sim 1.5$). This indicates that the film is much cleaner than the meander where the latter was eventually patterned by FIB milling using Ga ions which may introduce Ga contamination to the sample [28].

In the following sections, we present the low temperature transport characteristics for the meander device in detail. The linear scale representation of field dependent $R(T)$ data is displayed in Fig 2(b). For zero-field, the device offers mainly a three-step wide NM-SC transition and the temperature points $T_1$ (7.1 K), $T_2$ (6.9 K) and $T_3$ (6.2 K), shown by the dotted vertical lines, reflect the above-mentioned three-steps. Here, the first two transition steps appear sharp compared to that of the third one which shows a wide and slowly varying resistive tail type of characteristics at temperature below $T_3$. As the NM-to-SC transition is wide, $T_c$ is defined as the temperature where resistance drops to 50% of $R_N$. Accordingly, the $T_c$ obtained from zero-field $R(T)$ is 6.63 K which is shown by the violet dotted vertical line in Fig. 2(b). The dashed vertical line represents the transition temperature corresponding to the film, $T_c^{film}$ which is about 8.8 K.

Under external magnetic field, the steps in the transition region are clearly evident up to 250 mT. At 500 mT, the first two steps become smooth and transform into a continuously varying $R(T)$, however, the resistive tailing corresponding to the third step remains prominent. Under relatively high field of about 2.5 T, only a little drop in resistance occurs and with further increasing field, the resistance remains almost unaltered. However, in order to have a closer look onto the high field *R(T)* data, a selected region from Fig. 2(b) as bounded by the dotted rectangle is magnified in Fig. 2(c). We observe a weak dependence of resistance on the temperature at field ≥ 5T and the *R(T)* data shows a slight upturn indicating the onset of insulator-type behavior at high field [30]. The dashed black vertical line in Fig. 2(c) relates to the transition temperature of the film ($T_c^{film}$) where the drop in resistance of about 10 Ω is consistent with the film resistance as presented in Fig 2(a).

A semi-logarithmic presentation of the transition region from the zero-field *R(T)* for the meander is presented in Fig. 2(d). Below $T_c$, a majority of the transition region appears as resistive tailing. In general, wide transition width, tailing in the resistance and presence of residual resistance at *T<$T_c$* are the common features of PS events [7]. At *T ≈ $T_c$*, the PS excitations are thermally activated over an energy barrier ΔF and the corresponding TAPS contribution to the resistance can be described by the theory of Langer-Ambegaokar-McCumber-Halperin (LAMH) [30],

$$R_{TAPS}(T) = \frac{\pi \hbar^2 \Omega}{2e^2 k_B T} exp\left(\frac{-\Delta F(T)}{k_B T}\right) \quad (1)$$

Where, the attempt frequency, $\Omega = \frac{L}{\xi(T)} \left(\frac{\Delta F(T)}{k_B T}\right)^{1/2} \frac{1}{\tau_{GL}}$, with Ginzburg-Landau (GL) relaxation time, $\tau_{GL} = [\pi \hbar / 8 k_B (T_C - T)]$, and the coherence length, $\xi(T) = \xi(0)(1 - T/T_C)^{-1/2}$.

The energy barrier,

$\Delta F(T) = (8\sqrt{2}/3)(H_c^2(T)/8\pi)S\xi(T) \approx 0.83 \left(\frac{L}{\xi(0)}\right)\left(\frac{R_Q}{R_N}\right) k_B T_C \left(1 - \frac{T}{T_C}\right)^{3/2}$ where, $H_c$ is the critical field, *S* is the cross section of the nanowire, the quantum resistance, $R_Q = h/4e^2 = 6.45 k\Omega$ and $R_N$ is the normal state resistance.

Therefore,

$$R_{TAPS} = A'_1 t^{-3/2}(1-t)^{9/4} exp[-A'_2(1-t)^{3/2}t^{-1}] \quad (2)$$

Where, the reduced temperature $t = \frac{T}{T_c}$, $A'_2 = 0.83 \left(\frac{L}{\xi(0)}\right)\left(\frac{R_Q}{R_N}\right)$, and $A'_1 = (8/\pi)\left(\frac{L}{\xi(0)}\right)R_Q\sqrt{A'_2}$,

The PS model is usually applicable for 1D nanowire. In the present study, the $\xi$ for the device is about 9-10 nm [Fig. S2 in Supplementary Information (SI)] whereas, the thickness and the width of the meander are 50 nm and 200 nm, respectively. However, 1D model was extended to quasi 1D macroscopic crystals [19, 31, 32], 2D superconducting strips [12, 33], wide percolating films [34], nano-constrictions [35-37] nanowire network [38, 39] etc. Further, for FIB fabricated nanowires, due to Ga deposition and possible oxidation from the sidewall, the effective width becomes much smaller than the measured width [28, 29, 40]. Here, the red curve in Fig. 2(d) is the TAPS fit using Eqn. 2 where $\frac{L}{\xi(0)}$ and the $T_c$ are used as free parameters. The best fit is obtained for the $T_c$ value $T_c^{fit} = (6.1 \pm 0.24)\ K$ & $\frac{L}{\xi(0)} = 3.51 \pm 1.16$. Here, $R_N = 460\ \Omega$ is used to fit the data in Eqn. 2.

The multi-step transition might appear from the current crowding at the corner of the meander and/or due to the variation in width at the rounding part and at the lines of the meander [41]. Further, the broad and smoothly varying transition region below $T_3$ may relate to the weakest segment where the phase fluctuation is maximum as supported by the TAPS fit in this regime. The value of $\frac{L}{\xi(0)}$ from the fit also indicates a smaller segment with an effective length of about 3.5 $\xi$ to contribute to the PS events. This is supported by another device with a clearly visible nano-constriction where TAPS fit leads to $\frac{L}{\xi(0)} < 1$ (Fig. S1 in the SI). Further, the $T_c^{fit}$ is very close to the third transition point $T_3$ (6.2 K), hence the broadening in the third transition regime is mainly dominated by PS effects. Here, the normal state resistance $R_N^{fit}$ is fixed at 460 Ω as the transition starts from 460 Ω to down in the region fitted with Eqn. 2.

The PS process can also be supported by the appearance of intermediate resistive steps (IRS) in the current-voltage characteristics (IVCs) as presented in Fig. 3 for zero external magnetic field. Previously, Nb thin-films having thickness in the range of 8 - 50 nm are shown to exhibit single step sharp transition in their IVCs [8, 15]. However, the meander device in Fig. 3 clearly showcases different nature of IVCs featuring multiple IRS in the transition region. Fig. 3(a) represents a set of IVC isotherms measured in the temperature range 2.0 – 2.5 K for both up and down current sweep directions as shown by the black solid arrows. The critical current ($I_{c0}$) and the retrapping current ($I_{r1}$), marked by the dotted arrows, are different with the former exceeding the latter. Further, for the up-sweep direction, the SC-to-NM transition moves towards lower current for increasing temperature, whereas, in the down-sweep direction for NM-to-SC transition, the IVCs merge on top of each other and the retrapping current remains almost unchanged. Hence, the IVCs are hysteretic in nature and the hysteresis decreases with increasing temperature. The hysteresis is very common in granular films as already has been reported and it is related to Joule heating [8, 15, 42-44]. Also, a recent study on imaginary impedance measurements shows an enhancement of the same at higher currents which might lead to additional heating and hence an increment in the effective temperature at the sample end [45]. Further, the IVCs in the temperature range 2.0 - 2.3 K appear sharp and nearly single step SC-to-NM transition whereas, at 2.5 K, IVC looks noticeably different by featuring staircase type of multiple IRS. Note, the down sweeps are not like the up sweeps but they also feature IRS. The IVC measured at 2.5 K is presented separately in the inset of Fig. 3(a) where the IRS are observed to merge at the excess current, $I_s$ as shown by the dotted cyan lines. Further the IRS feature increasing slope in the increased current direction. Hence, the origin of the IRS is likely to be related to PS [9, 24].

The evolution of IVC isotherms for the temperature range 2.5 - 4.9 K is shown in Fig. 3(b). The solid and open circles represent the up and down sweep directions, respectively. For both the sweep directions, transition moves towards lower current with increasing temperature. However, the shift is more for the up sweep than that of the down sweep. The IRS are reproducible and the slopes for the resistive steps align with each other among the individual IVCs. Clearly from Fig. 3(b), there are two major resistive slopes that are followed by all the IVCs in between the SC and NM states. Here, as shown by the black arrows, the onset of finite voltage from the zero-voltage level and those two consecutive major resistive slopes are marked as the critical currents $I_{c0}$, $I_{c1}$ and $I_{c2}$, respectively. The black dotted rectangular box surrounding these three characteristic critical currents is highlighted in Fig 3(c) where the formation of minor sub-resistive steps, shown by the dotted circular region in between $I_{c0}$ and $I_{c1}$ at temperature $\geq$ 3.5 K, is evident. Finally, above 4.5 K, the branch related to $I_{c1}$ disappears and the sub-resistive steps merge onto $I_{c2}$. Here, with increasing temperature major resistive steps split into several sub resistive steps before they merge onto the next major step. To summarize, a few selective IVC isotherms in both sweep directions are presented in Fig. 3(d). The hysteresis in the IVCs decreases with increasing temperature and the number of resistive steps decreases at the temperatures close to the $T_c$. Finally, a complete metallic behavior by the linear IVC at 7.1 K indicates that the $T_c$ from IVC matches with the transition temperature $T_1$ as shown in Fig 2(b).

The magnetic field-dependent IVC isotherms for three selective fields are presented in Fig. 4 for up sweep direction only. The field was applied perpendicular to the sample plane. Fig 4(a) represents the IVCs measured under 75 mT field and in the temperature range 2.0 -7.1 K. Unlike the zero-field IVCs [Fig. 3(a)], IRS start to appear here from the base temperature 2.0 K and the rest of the variations remained similar to that of the zero-field IVCs. The major steps bounded by the black dotted rectangular box is magnified in Fig. 4(b). The $I_{c0}$, $I_{c1}$ and $I_{c2}$ are marked by the black arrows. At 3.25 K, the sub-resistive minor steps start to appear and they merge onto the slope $I_{c1}$. At temperature $\geq$ 4.5 K, IVCs follow the slope of $I_{c2}$ only and therefore, 4.5 K can be considered as a crossover point where the clear separation of the two slopes is visible. Before merging onto $I_{c2}$, minor steps are visible in Fig. 4(b) for the IVCs measured at 4.5 K and above. In a similar fashion, IVC isotherms for 250 mT and 500 mT have been shown in (c) and (e), respectively. The regions covering the three characteristic currents in (c) & (e) are zoomed in (d) & (f), respectively. For 250 mT, IVCs look almost similar to that with 75 mT, except for the appearance of sub-resistive minor steps at 2K and a reduced $I_{c0}$ from 59.38 µA for 75 mT to 31.25 µA for 250 mT at the base temperature. The transit temperature which separates the two major steps related of $I_{c1}$ and $I_{c2}$ is 4.3 K for 250 mT. For the increased field of 500 mT, $I_{c0}$ reduces further and it is about 25 µA at 2.0 K as seen in Fig. 4(e) & (f). Here, overall IVCs look smoother with lesser number of visible steps. However, there is no clear separation between the slopes $I_{c1}$ and $I_{c2}$ as it appeared in the previous cases with lower field values. Interestingly, the field-dependent *R(T)* data presented in Fig. 2(b) also showed a relatively smooth transition for 500 mT compared to that for field $\leq$ 250 mT which consisted of multiple kinks in the *R(T)* transition.

Fig. 5 presents the effect of magnetic field on the IVCs in a different way where the IVCs are collected from different magnetic fields while keeping the temperature fixed. We have selected two specific temperatures, viz, the base temperature 2 K and 2.5 K. At 2.5 K, IRS start to show up for the zero-field IVCs as shown in Fig. 3(a). The field dependent IVCs measured at 2 K for both sweep directions have been displayed in Fig. 5(a). Here, the field was applied up to 1T. Similar to the zero-field IVCs [Fig. 3 (a)], at lower field values such as 25 mT, a single step sharp transition is observed. However, with a slight increase in field up to 50 mT, the transition no longer remains a single-step transition but it starts to accompany IRS as seen by the green curve in Fig. 5(a). The steps become more prominent for increased fields. The hysteresis, originated by the differences in up and down sweep directions, reduces with increasing magnetic field and at about 1T, it disappears. The effect of the magnetic field is similar to that of the temperature. For clarity, in Fig. 5(b), we have selectively presented IVCs measured under field $\geq$ 50 mT for up sweep only. The extended red dotted lines, aligned with the IRS, converge at the excess current $I_s$ on the current-axis with increasing slopes in the increasing current direction. This indicates that the resistive states, appeared under external magnetic field, are originated from PS events [9, 24]. Furthermore, the major steps as bounded by the black dotted rectangular region in Fig. 5(b) is highlighted in Fig. 5(c) where the previously defined three characteristic critical currents $I_{c0}$, $I_{c1}$ and $I_{c2}$ are marked. The steps are wide and clear for fields $\geq$ 75 mT and at 250 mT, sub-resistive minor steps start to appear at 2.0 K. Further at 500 mT, the prominent slope follows $I_{c2}$ and the separation between $I_{c1}$ and $I_{c2}$ slopes is evident while increasing the field from 250 mT to 500 mT. At 1T, the steps become smoother compared to that from the lower fields.

Similarly, IVCs measured at 2.5 K under various magnetic field are shown in Fig. 5(d) where the main panel displays both up and down sweeps and the inset shows a selected part from the up sweep only. As seen already, at 2.5 K, the zero-field IVC features IRS that become pronounced at higher magnetic fields. The characteristic critical currents are shown in the inset which shows the onset of sub-resistive minor steps at 250 mT before merging onto the slope $I_{c1}$. Here, the minor steps at 250 mT are wider and more prominent compared to that shown in Fig. 5(c) for 2.0 K.

We have extracted various characteristic currents from the IVCs presented in Figs. 3, 4 & 5 and their temperature dependence under external magnetic field is displayed in Fig. 6. Eventually, the characteristic currents are plotted against the reduced temperature ($T/T_c$) and the zero-field case is presented in Fig 6(a). Here, $T_c$ refers to the temperature at which IVC becomes linear. The characteristic currents are defined in the inset which presents a representative IVC measured at 2.7 K. For up sweep, four characteristic critical currents, namely, $I_{c0}$, $I_{c1}$, $I_{c2}$, $I_c^{nm}$ are considered and for the down sweep, two retrapping currents, $I_{r1}$ & $I_{r0}$ are measured. The critical currents $I_{c0}$, $I_{c1}$ and $I_{c2}$ are already defined in the previous section. Here, $I_c^{nm}$ is the current at which the device switches to the complete normal state. For the down sweep, $I_{r1}$ and $I_{r0}$ represent the currents at which voltage starts to drop and voltage drops to zero to reach the complete superconducting state, respectively. All these characteristic currents are marked by black arrows in the inset of Fig. 6(a).

For the temperatures below the red dotted vertical line in the main panel of Fig. 6(a), critical currents $I_{c0}$ and $I_c^{nm}$ merge with each other indicating a single step SC-to-NM transition. Further, $I_{c1}$ and $I_{c2}$ start only from the same temperature point as indicated by the red dotted vertical line which thus acts as the cross over temperature from a single step to a multistep transition.

Besides, $I_{c0}$ varies non-trivially with temperature by showcasing random oscillatory behavior before merging with $I_{r0}$. However, there are very little variations in $I_{c1}$, $I_{c2}$ and $I_c^{nm}$ with temperature for T < $0.6T_c$, at which $I_{c1}$ starts to drop drastically and at T= $0.66T_c$ (~ 4.7 K with $T_c$ = 7.1 K), it reduces from ~ 58 µA to ~ 17 µA and finally it disappears at $0.77T_c$. Contrary to that, $I_{c2}$ remains prominent up to $0.9T_c$. As already observed in Fig. 3(c) that at about 4.7 K, the major resistive slope related to $I_{c1}$ disappears and the resistive steps merge mainly with the slope related to $I_{c2}$. However, the values of $I_{c1}$ in the temperature range $0.66T_c$ < T < $0.77T_c$ have been obtained from the sub-resistive steps that follow the same slope related to $I_{c1}$. However, the retrapping currents $I_{r0}$ & $I_{r1}$ vary similarly with temperature except for a wide gap between them indicating a broad wide transition width in the returning down sweep also. Here, $I_{r0}$ remains smaller than the remaining characteristic currents before merging with the latter. But, $I_{r1}$ exceeds $I_{c0}$ at the red dotted vertical line and remains higher with increasing temperature. Nevertheless, $I_{r1}$ stays under the $I_c^{nm}$ value up to $0.92T_c$, but crosses $I_{c1}$ and $I_{c2}$ at temperatures between $0.6T_c$ and $0.7T_c$, respectively. The temperature at the crossing point of the retrapping current with the critical current is known as the threshold temperature at which point retrapping current exceeds the critical current[42]. Finally, only one critical current exists at T<$0.35T_c$ corresponding to the red dotted vertical line, indicating that at T > $0.35T_c$, phase slips start to show up by featuring IRS in the IVCs.

Temperature dependence of the critical and retrapping currents under 75 mT, 250 mT and 500 mT field are presented in Fig 6 (b), (c) and (d), respectively. Overall values for these characteristic currents decrease with increasing field. Importantly, $I_{c0}$ and $I_c^{nm}$ are having different values and the resistive steps start from the base temperature for all the three magnetic fields. Therefore, PS effects get facilitated by the application of magnetic field as reported previously [15, 16]. $I_{c0}$ varies with temperature in a similar fluctuating trend for 75 mT as that of zero-field case, however, for higher field, the fluctuations smoothen up. Further, $I_{r1}$ starts to exceed the critical currents $I_{c1}$, $I_{c2}$, $I_c^{nm}$ at lower temperature with increasing field and eventually, it reaches above of all the critical currents for 500 mT. For 250 mT, at T<0.5$T_c$, $I_{c2}$ and $I_c^{nm}$ are the same and at higher temperature they split and separate values are obtained as shown in Fig. 6(c). Sudden drop in $I_{c1}$ is also evident for the cases of 75 mT and 250 mT. Finally, the observations from the IVCs presented in the previous figures are clearly reflected in Fig. 6.

Now, we consider $I_{c0}$ in greater detail as it refers to the current at which a finite voltage starts to appear and thus $I_{c0}$ is the maximum dissipation-less current the device can carry. In Fig. 7, we present the variation of $I_{c0}$ with respect to magnetic field and the measurement temperature. First, the variation of $I_{c0}$ with magnetic field for 2.0 K, 2.1 K, 2.2 K, 2.4 K and 2.7K is shown in Fig 7(a). The critical current strongly depends on temperature for magnetic field < 100 mT. However, above 100 mT, $I_{c0}$ does not vary so significantly. Here, little increment in temperature reduces the critical current and at 2.7 K, $I_{c0}$ remains almost flat even at the lower field regime as shown by the magnified plot in the inset of Fig. 7(a). The strongest dependence for $I_{c0}$ on the field occurs at the base temperature 2.0 K which shows a steep change in $I_{c0}$ with the reduction in magnetic field in the low field regime up to ~ 100 mT.

Further, to get an idea about the temperature range, the variation of $I_{c0}$ with temperature for specific magnetic field is shown in Fig. 7(b). It is clear that in the low field range, the $I_{c0}$ varies prominently for temperature up to ~ 3.0 K. At fields ≥ 250 mT, $I_{c0}$ varies smoothly. The low temperature region as bounded by the dotted rectangular box in Fig. 7(b) is highlighted for better view in Fig. 7(c) where it is evident that $I_{c0}$ changes significantly at low temperature and particularly for low field values. In order to have a better comparison of $I_{c0}$ with temperature and magnetic field at one single platform, in Fig. 7 (d), we have plotted $I_{c0}$ with respect to the reduced temperature for the magnetic fields up to 500 mT in 3D representation. The amplitude as well as the fluctuating nature of the variation of $I_{c0}$ particularly for low temperature ($< 0.5T_c$) and low magnetic field values ($\leq 100$ mT) are apparent. With increasing temperature and increased magnetic field $I_{c0}$ becomes smooth and variation is not so large in amplitude.

## 4. Conclusions:

In conclusion, Nb thin films, patterned into meander structures by FIB milling, have been studied by low temperature transport measurements. Comparing with the original thin-film device, meandering leads to a few orders of increment in the normal state resistance and hence, the device's overall electrical characteristics are governed mainly by the meander structure. By meandering, the SC-to-NM transition, as appeared in $R(T)$ measurements, transforms from a sharp to a wide and multistep transition. Further, below the $T_c$, $R(T)$ features resistive tailing which is explained by the TAPS mechanism based on LAMH model and the TAPS fit indicates towards a selected segment of the meander to contribute majorly to the PS process. Further, the SC-NM transition have been investigated in detail through IVC measurements with and without the presence of magnetic field. Interestingly, the IVCs present wide transition too while featuring staircase like intermediate resistive steps that confirm the PS events to take place in the meander structure [9, 24, 29]. Here, magnetic field and the temperature are shown to facilitate the PS process in a similar fashion. Furthermore, particularly for temperatures $< 3.0$ K and field $< 100$ mT, the critical current strongly varies with temperature and magnetic field. It shows a non-trivial oscillatory type fluctuating behaviour which smears out at higher temperature and higher field. Finally, the results are clearly indicative of the PS process occurring due to meandering the thin-film. In future, it would be interesting to monitor phase slip nucleation process during the measurement of IVCs by means of imaginary impedance measurements that can probe the thermal variations more accurately [45]. Further, in order to have more insight into the fluctuating behaviour of the measured critical current at low temperature and its possible application in sensor-based devices, a detailed study with precise variations in magnetic field at low field range is needed and will be carried out in near future.

**Data availability**

The data that represent the results in this paper and the data that support the findings of this study are available from the corresponding author upon reasonable request.

**Acknowledgements**

We are thankful to Ms. Ambika Bawa and Mr. M. B. Chhetri for their assistance in building the sputtering unit at the initial stages. D.S. and B. G. acknowledges to UGC-RGNF for providing SRF. S.Y. acknowledges UGC for his SRF. This work was supported by CSIR network project 'AQuaRIUS' (Project No. PSC 0110) and was carried out under the mission mode project "Quantum Current Metrology".

**ORCID iDs:**

Sangeeta Sahoo https://orcid.org/0000-0002-7189-4561

**References**

[1] Arutyunov KY, Golubev DS, Zaikin AD 2008 Superconductivity in one dimension *Phys. Rep.* **464** 1-70

[2] Giordano N 1988 Evidence for Macroscopic Quantum Tunneling in One-Dimensional Superconductors *Phys. Rev. Lett.* **61** 2137-40

[3] Kato K, Takagi T, Tanabe T, Moriyama S, Morita Y, Maki H 2020 Manipulation of phase slips in carbon-nanotube-templated niobium-nitride superconducting nanowires under microwave radiation Sci. Rep. **10** 14278

[4] Mooij JE, Nazarov YV 2006 Superconducting nanowires as quantum phase-slip junctions *Nat. Phys.* **2** 169-72

[5] Berdiyorov GR, Milošević MV, Peeters FM 2009 Kinematic vortex-antivortex lines in strongly driven superconducting stripes *Phys.Rev. B* **79** 184506

[6] Baranov VV, Balanov AG, Kabanov VV 2013 Dynamics of resistive state in thin superconducting channels *Phys. Rev. B* **87** 174516

[7] Zhao W, Liu X, Chan MHW 2016 Quantum Phase Slips in 6 mm Long Niobium Nanowire. *Nano Lett.* **16** 1173-8

[8] Bawa A, Jha R, Sahoo S 2015 Tailoring phase slip events through magnetic doping in superconductor-ferromagnet composite films *Sci. Rep.* **5** 13459

[9] Sivakov AG, Glukhov AM, Omelyanchouk AN, Koval Y, Müller P, Ustinov AV 2003 Josephson Behavior of Phase-Slip Lines in Wide Superconducting Strips. *Phys. Rev. Lett.* **91** 267001

[10] Falk A, Deshmukh MM, Prieto AL, Urban JJ, Jonas A, Park H 2007 Magnetic switching of phase-slip dissipation in NbSe$_2$ nanoribbons *Phys. Rev. B* **75** 020501

[11] Kimmel G, Glatz A, Aranson IS 2017 Phase slips in superconducting weak links *Phys. Rev. B* **95** 014518

[12] Qiu C, Qian T 2008 Numerical study of the phase slip in two-dimensional superconducting strips *Phys. Rev. B* **77** 174517

[13] Córdoba R, Baturina TI, Sesé J, Mironov AY, Teresa JMD, Ibarra MR, Nasimov DA, Gutakovskii AK, Latyshev AV, Guillamón I, Suderow H, Vieira S, Baklanov MR, Palacios JJ, Vinokur VM 2013 Magnetic field-induced dissipation-free state in superconducting nanostructures *Nat. Commun.* **4** 1437

[14] Astafiev OV, Ioffe LB, Kafanov S, Pashkin YA, Arutyunov KY, Shahar D 2012 Coherent quantum phase slip *Nature* **484** 355-8

[15] Gajar B, Yadav S, Sawle D, Maurya KK, Gupta A, Aloysius RP 2019 Substrate mediated nitridation of niobium into superconducting Nb2N thin films for phase slip study *Sci. Rep.* **9** 8811

[16] Yadav S, Kaushik V, Saravanan MP, Aloysius RP, Ganesan V, Sahoo S 2021 A robust nitridation technique for fabrication of disordered superconducting TiN thin films featuring phase slip events. *Sci. Rep.* **11** 7888

[17] Bawa A, Gupta A, Singh S, Awana VPS, Sahoo S 2016 Ultrasensitive interplay between ferromagnetism and superconductivity in NbGd composite thin films *Sci. Rep.* **6** 18689

[18] Lehtinen JS, Sajavaara T, Arutyunov KY, Presnjakov MY, Vasiliev AL 2012 Evidence of quantum phase slip effect in titanium nanowires *Phys. Rev. B* **85** 094508

[19] Petrović AP, Ansermet D, Chernyshov D, Hoesch M, Salloum D, Gougeon P, et al 2016 A disorder-enhanced quasi-one-dimensional superconductor *Nat. Commun.* **7** 12262

[20] Constantino NGN, Anwar MS, Kennedy OW, Dang M, Warburton PA, Fenton JC 2018 Emergence of Quantum Phase-Slip Behaviour in Superconducting NbN Nanowires: DC Electrical Transport and Fabrication Technologies *Nanomaterials* **8** 442.

[21] Constantino NGN 2016 Disorder in Superconductors in Reduced Dimensions [Ph.D. Thesis]: University College London, Gower Street, London, UK

[22] Bartolf H, Engel A, Schilling A, Il'in K, Siegel M, Hübers HW, et al 2010 Current-assisted thermally activated flux liberation in ultrathin nanopatterned NbN superconducting meander structures *Phys. Rev. B* **81** 024502

[23] Joshi LM, Rout PK, Husale S, Gupta A 2020 Dissipation processes in superconducting NbN nanostructures *AIP Adv.* **10** 115116

[24] Delacour C, Pannetier B, Villegier J-C, Bouchiat V 2012 Quantum and Thermal Phase Slips in Superconducting Niobium Nitride (NbN) Ultrathin Crystalline Nanowire: Application to Single Photon Detection *Nano Lett.* **12** 3501-6

[25] Aloysius RP, Husale S, Kumar A, Ahmad F, Gangwar AK, Papanai GS, et al 2019 Superconducting properties of tungsten nanowires fabricated using focussed ion beam technique *Nanotechnology* **30** 405001

[26] Nazir M, Yang X, Tian H, Song P, Wang Z, Xiang Z, et al 2020 Investigation of dimensionality in superconducting NbN thin film samples with different thicknesses and NbTiN meander nanowire samples by measuring the upper critical field *Chin. Phys. B* **29** 087401

[27] Zhao L, Jin Y-R, Li J, Deng H, Zheng D-N 2014 Fabrication and properties of the meander nanowires based on ultra-thin Nb films *Chin. Phys. B* **23** 087402

[28] Tettamanzi GC, Pakes CI, Potenza A, Rubanov S, Marrows CH, Prawer S 2009 Superconducting transition in Nb nanowires fabricated using focused ion beam. *Nanotechnology* **20** 465302

[29] Lyatti M, Wolff MA, Savenko A, Kruth M, Ferrari S, Poppe U, et al 2018 Experimental evidence for hotspot and phase-slip mechanisms of voltage switching in ultrathin $YBa_2Cu_3O_{7-x}$ nanowires *Phys. Rev. B* **98** 054505

[30] Bezryadin A 2008 Quantum suppression of superconductivity in nanowires *J. Phys. Condens. Matter* **20** 043202

[31] Mitra S, Petrović AP, Salloum D, Gougeon P, Potel M, Zhu J-X, et al 2018 Dimensional crossover in the quasi-one-dimensional superconductor $Tl_2Mo_6Se_6$ *Phys. Rev. B* **98** 054507

[32] Ansermet D, Petrović AP, He S, Chernyshov D, Hoesch M, Salloum D, et al 2016 Reentrant Phase Coherence in Superconducting Nanowire Composites *ACS Nano* **10** 515-23

[33] Bell M, Sergeev A, Mitin V, Bird J, Verevkin A, Gol'tsman G 2007 One-dimensional resistive states in quasi-two-dimensional superconductors: Experiment and theory *Phys. Rev. B* **76** 094521

[30] Qiu C, Qian T 2008 Numerical study of the phase slip in two-dimensional superconducting strips *Phys. Rev. B* **77** 174517

[34] Amol N, Shawn F, Jack G, Alex S, Kristiaan T, Margriet JVB, et al 2017 Quantum fluctuations in percolating superconductors: an evolution with effective dimensionality *Nanotechnology* **28** 165704

[35] Baumans XDA, Cerbu D, Adami O-A, Zharinov VS, Verellen N, Papari G, et al 2016 Thermal and quantum depletion of superconductivity in narrow junctions created by controlled electromigration *Nat. Commun.* **7** 10560

[36] Voss JN, Schön Y, Wildermuth M, Dorer D, Cole JH, Rotzinger H, et al 2021 Eliminating Quantum Phase Slips in Superconducting Nanowires *ACS Nano* **15** 4108-14

[37] Kenawy A, Magnus W, Milošević MV, Sorée B 2020 Electronically tunable quantum phase slips in voltage-biased superconducting rings as a base for phase-slip flux qubits *Supercond. Sci. Technol.* **33** 125002

[38] Trezza M, Cirillo C, Sabatino P, Carapella G, Prischepa SL, Attanasio C 2013 Nonlinear current-voltage characteristics due to quantum tunneling of phase slips in superconducting Nb nanowire networks *Appl. Phys. Lett.* **103** 252601

[39] Cirillo C, Trezza M, Chiarella F, Vecchione A, Bondarenko VP, Prischepa SL, et al 2012 Quantum phase slips in superconducting Nb nanowire networks deposited on self-assembled Si templates *Appl. Phys. Lett.* **101** 172601

[40] Rogachev A, Wei TC, Pekker D, Bollinger AT, Goldbart PM, Bezryadin A 2006 Magnetic-Field Enhancement of Superconductivity in Ultranarrow Wires *Phys. Rev. Lett.* **97** 137001

[41] Engel A, Renema JJ, Il'in K, Semenov A 2015 Detection mechanism of superconducting nanowire single-photon detectors *Supercond. Sci. Technol.* **28** 114003

[42] Hazra D, Pascal LMA, Courtois H, Gupta AK 2010 Hysteresis in superconducting short weak links and micro-SQUIDs *Phys. Rev. B* **82** 184530

[43] Skocpol WJ, Beasley MR, Tinkham M 1974 Self-heating hotspots in superconducting thin-film microbridges *J. Appl. Phys.* **45** 4054-66

[44] Das Gupta K, Soman SS, Sambandamurthy G, Chandrasekhar N 2002 Critical currents and vortex-unbinding transitions in quench-condensed ultrathin films of bismuth and tin *Phys. Rev. B* **66** 144512

[45] Perconte D, Mañas-Valero S, Coronado E, Guillamón I, Suderow H 2020 Low-Frequency Imaginary Impedance at the Superconducting Transition of 2*H*-NbSe$_2$ *Phys. Rev. Appl.* **13** 054040

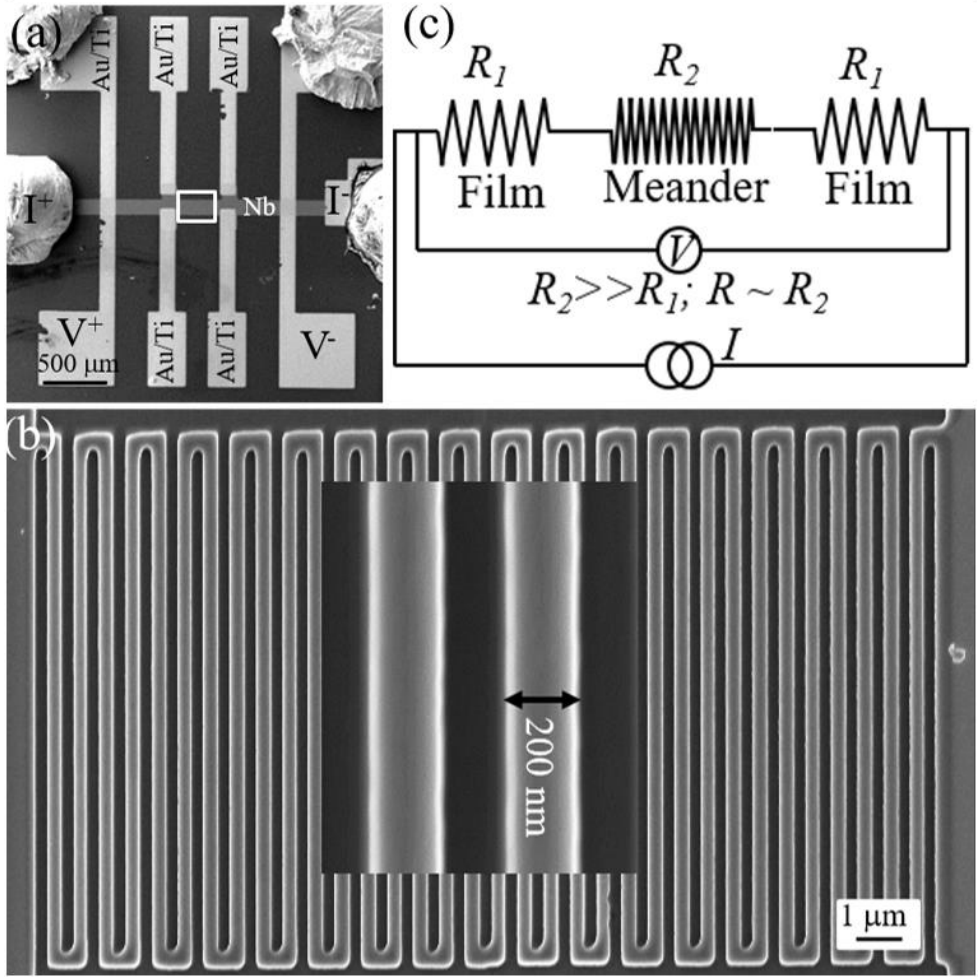

**Fig. 1:** Device geometry and equivalent circuit diagram. (a) Scanning electron microscopy (SEM) image of a device with Nb as the conducting channel and Au/Ti as electrical contacts for transport measurements. The rectangular box at the middle is patterned into a meander structure and the same is shown in (b). The individual meander line, its width and the gap between the lines are highlighted at the center of (b). (c) Equivalent circuit mainly based on two resistors, $R_1$ and $R_2$, representing the thin film and the meander structure, respectively. Here the latter dominates over the former as the resistance $R_2$ corresponding to the meander is much higher than the thin film resistance $R_1$.

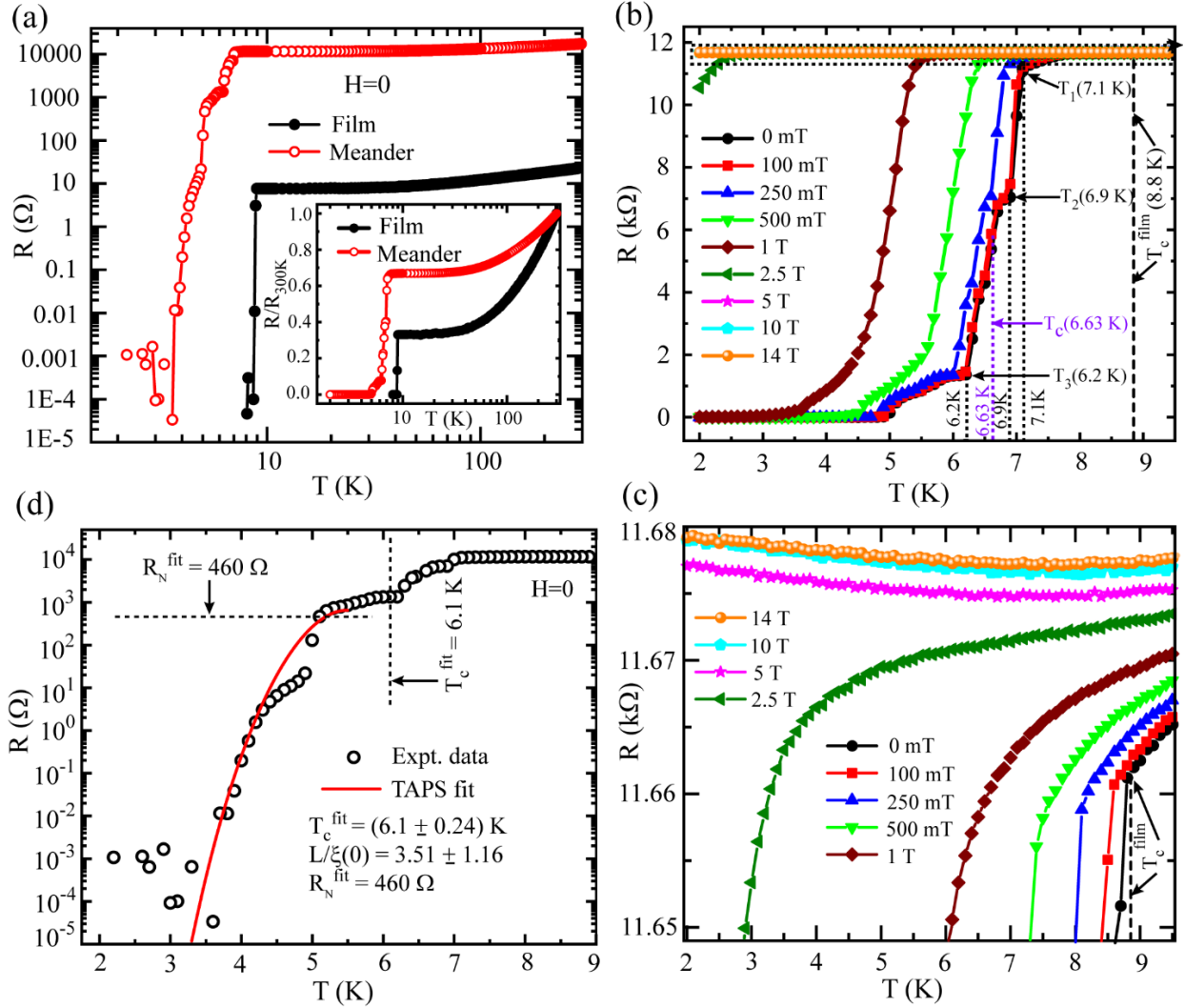

**Fig.2:** Temperature dependent resistance *R(T)* measurements with and without magnetic field. (a) *R(T)* comparison of thin film and meander by zero-field *R(T)* measurements before and after the meander fabrication. Inset: comparison of the normalized resistance for the film only and with the meander. Here, the resistance is normalized with respect to the resistance at 300 K for both the configurations, i.e., with and without meander. (b) *R(T)* measurements under external magnetic field applied perpendicular to the sample plane. The transition from normal state to the superconducting state occurs in three steps and the corresponding temperature values are marked by the dotted vertical lines. (c) Magnified view of a selected region close to the normal state as shown in (b). Here, the transition point corresponding to the film is clearly visible and is marked with the dotted vertical line. A slight upturn in *R(T)* for the field ≥ 5T is evident. (d) A semi-log representation of zero-field *R(T)* measured on the meander device for a selected range of temperature. The red curve corresponds to a fit based on the thermally activated phase slip (TAPS) model. The details are explained in the main text.

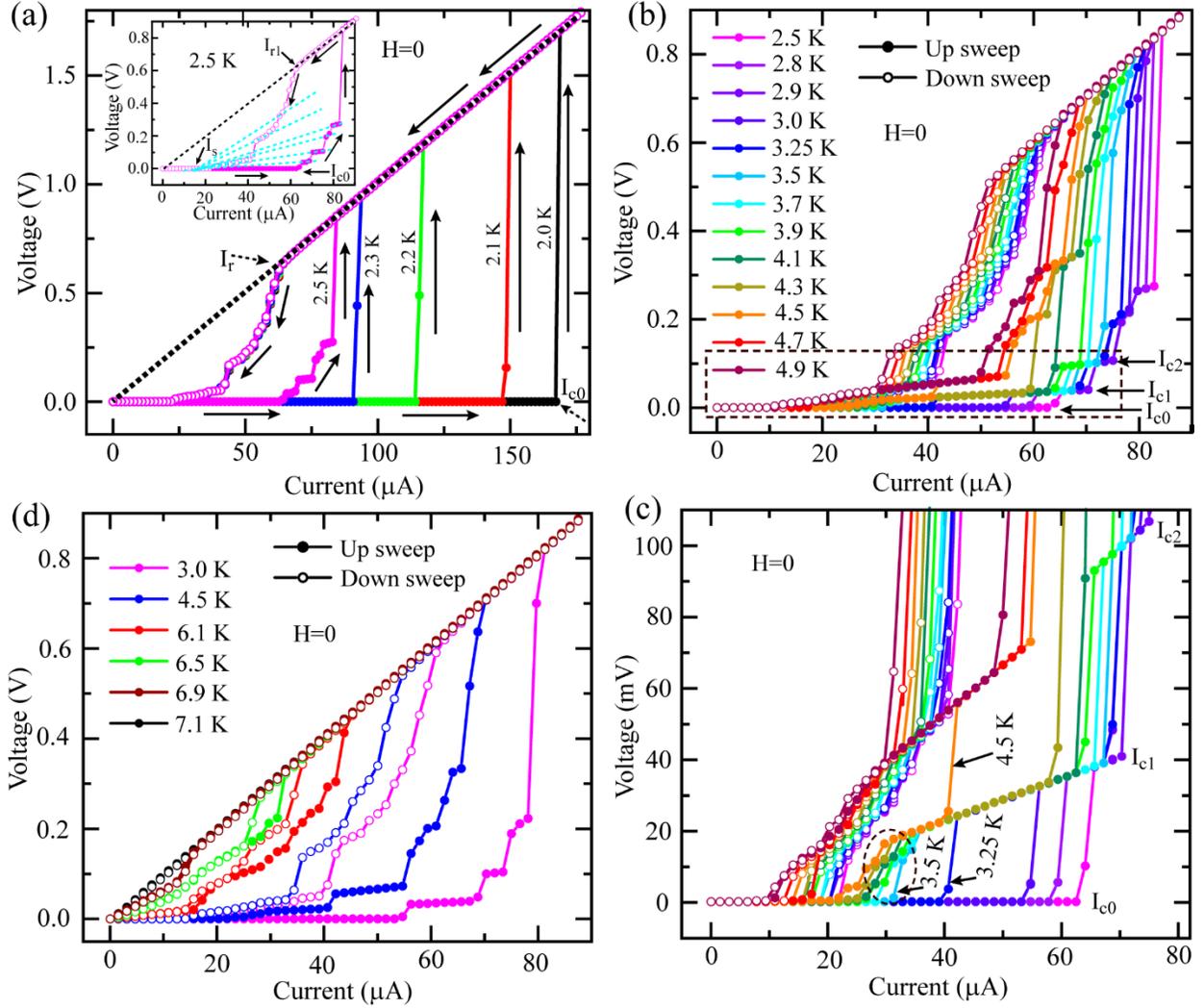

**Fig. 3:** Zero-field current voltage characteristics (IVCs) measured on the meander device. (a) A set of IVC isotherms measured in the temperature range 2.0 – 2.5 K. The black solid arrows indicate the sweeping direction with respect to the bias current. The critical current ($I_c$) and the re-trapping current ($I_r$) are marked by the dotted arrows. Inset: A separate presentation of a single IVC measured at 2.5 K where the intermediate resistive steps are observed to merge at the excess current, Is. (b) IVC isotherms measured at relatively higher temperature range from 2.5 K to 4.9 K. Here, as shown by the black arrows, the zero-voltage level and the two consecutive major resistive steps in the low voltage regime are marked as the critical currents $I_{c0}$, $I_{c1}$ and $I_{c2}$, respectively. The afore-mentioned critical currents bounded by the dotted rectangular box are highlighted in (c) which indicates the formation of minor sub-resistive steps in between $I_{c0}$ to $I_{c1}$ at temperature 3.5 K and above as shown by the dotted circular region. Finally, above 4.5 K the branch related to $I_{c1}$ disappears and the sub-resistive steps merge onto the slope related to $I_{c2}$. (d) A set of selective IVCs that reflect the change in hysteresis with the measurement temperature where the hysteresis is caused by the differences in up and down sweep direction of the bias current.

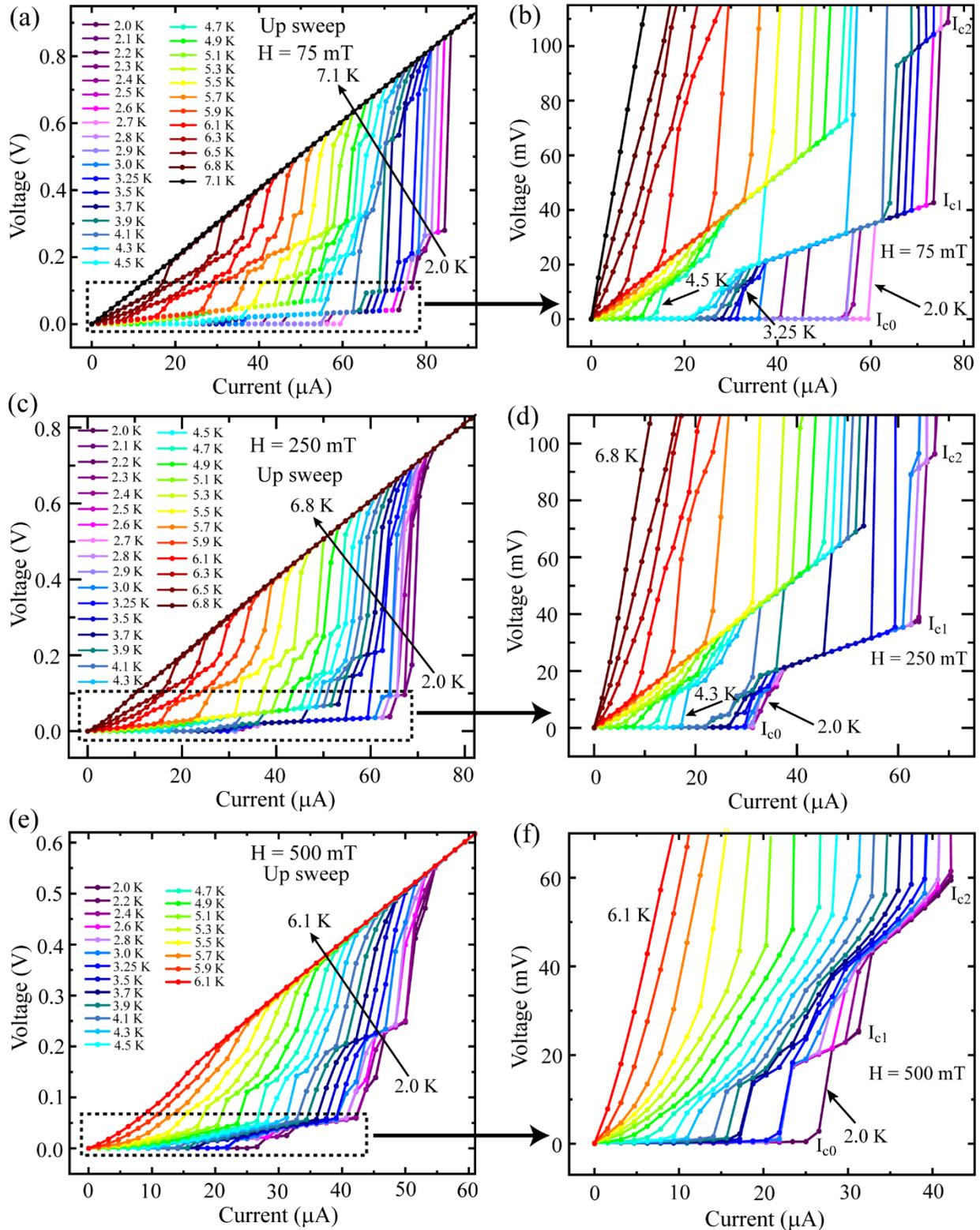

**Fig. 4:** Magnetic field-dependent IVC isotherms for only up sweep. Magnetic field was applied perpendicular to the sample plane. IVC isotherms measured under an applied magnetic field of (a) 75 mT, (c) 250 mT and (e) 500 mT. The dotted rectangular portions from (a), (c) and (e) are highlighted in (b), (d) and (f), respectively.

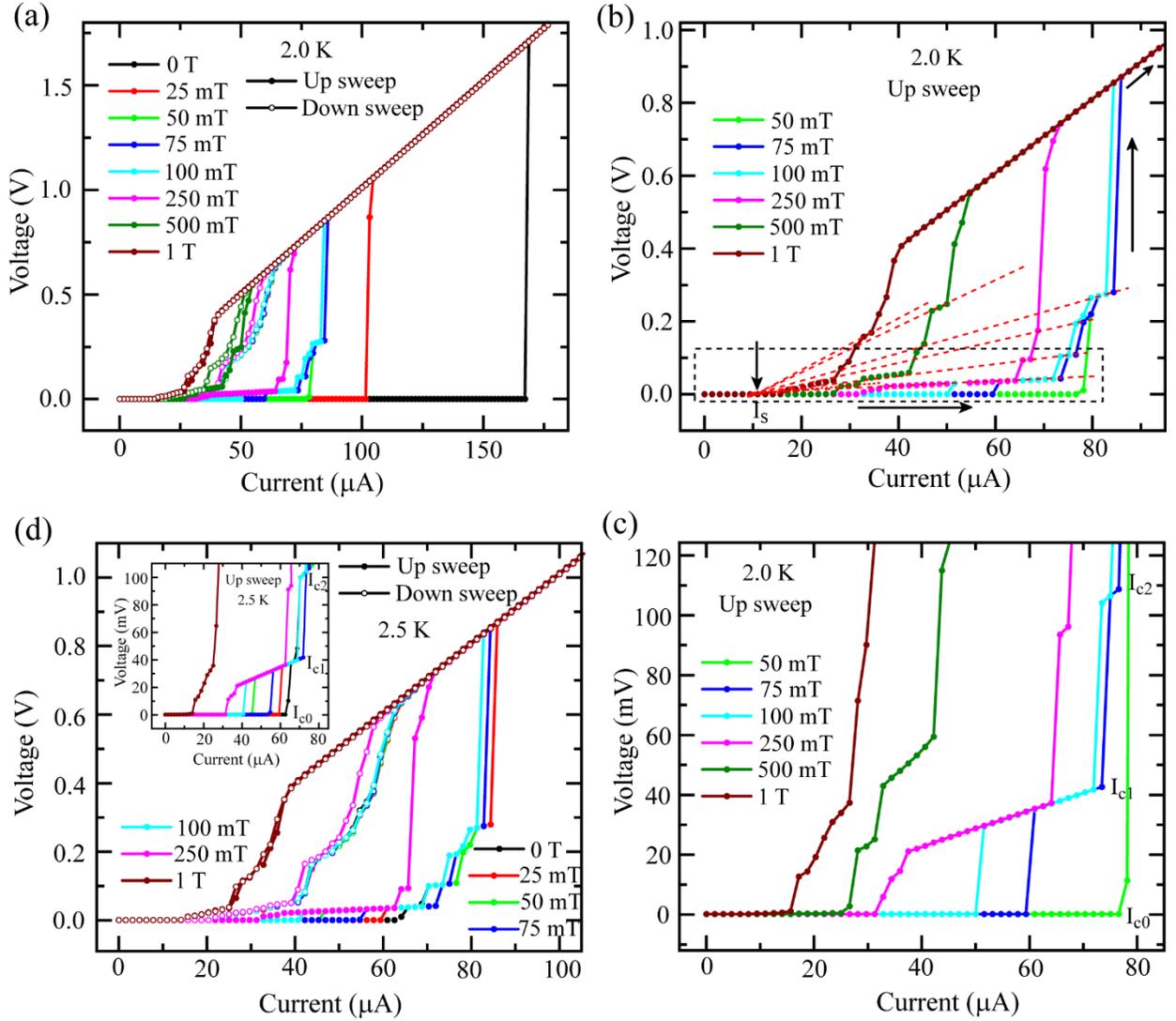

**Fig.5:** Variation of IVCs with magnetic field. Magnetic field dependent IVCs measured at 2.0 K (a) for both up and down sweeps and (b) for only Up sweeps for field ≥50 mT. The dotted red lines representing the slopes of the intermediate resistive steps merge at the excess current $I_s$. (c) A magnified view of the region bounded by the black dotted rectangle in (b). (d) Field dependent IVCs measured at 2.5 K for both up and down sweeps. Inset: low voltage region of the up sweep only consisting of the three characteristic currents $I_{c0}$, $I_{c1}$ and $I_{c2}$ for the IVC measured at 2.5 K.

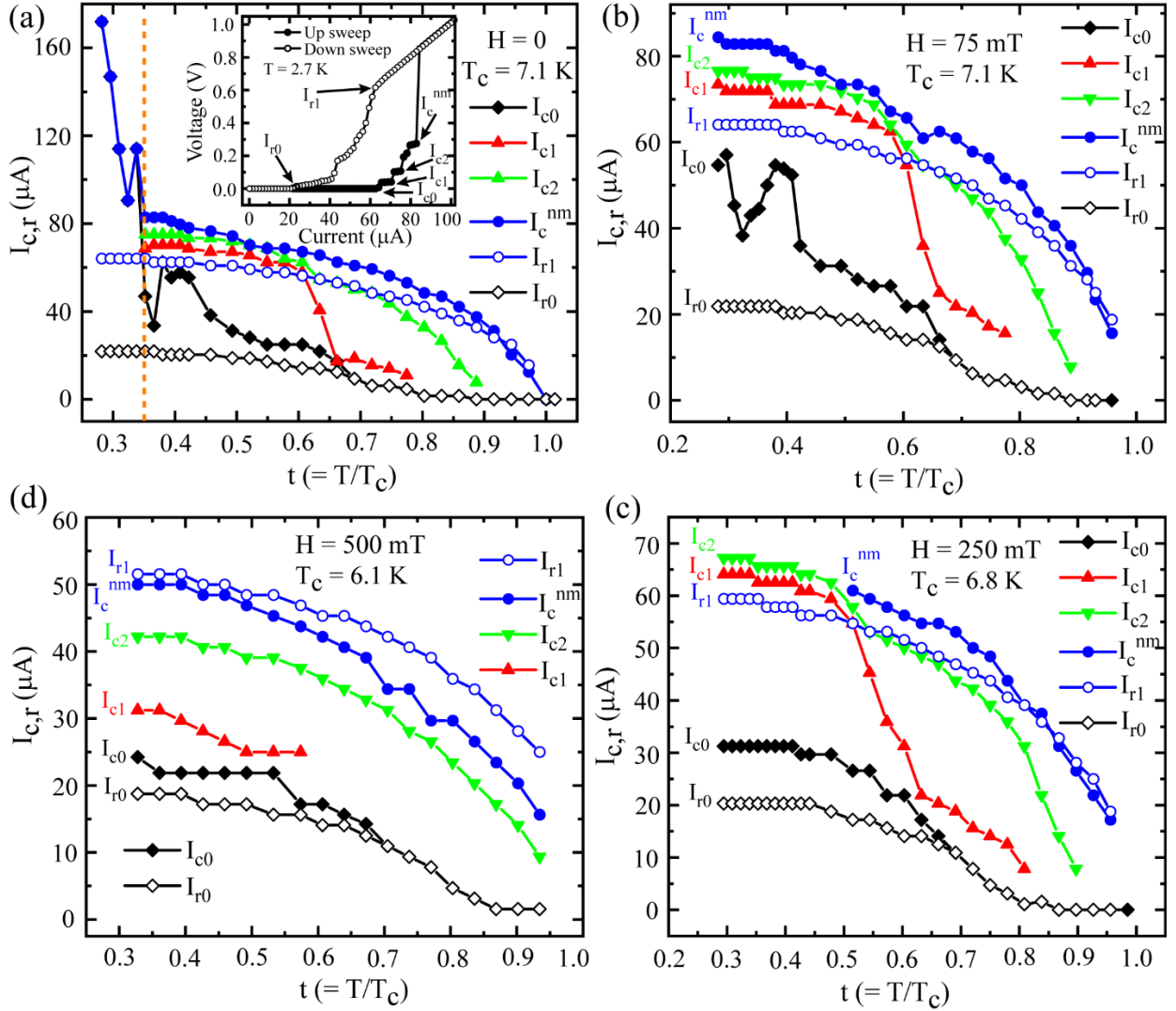

**Fig. 6:** Temperature dependence of characteristic currents and the effect of magnetic field. (a) the zero-field variations of the aforementioned characteristic currents on the reduced temperature ($T/T_c$). Where $T_c$ is obtained from the IVCs and the same is mentioned in the individual panel of the figure. The red dotted vertical line separates the single-step sharp transition from the multi-step wide transition featuring the phase slip events. Inset: the characteristic currents are defined for a representative IVC measured at 2.7 K for no external applied magnetic field. For up-sweep direction, the characteristic currents refer to various critical current and for down sweep direction, the same is indicated as retrapping current. The variation of the defined critical and retrapping currents with reduced temperature for (b) 75 mT, (c) 250 mT and (d) 500 mT.

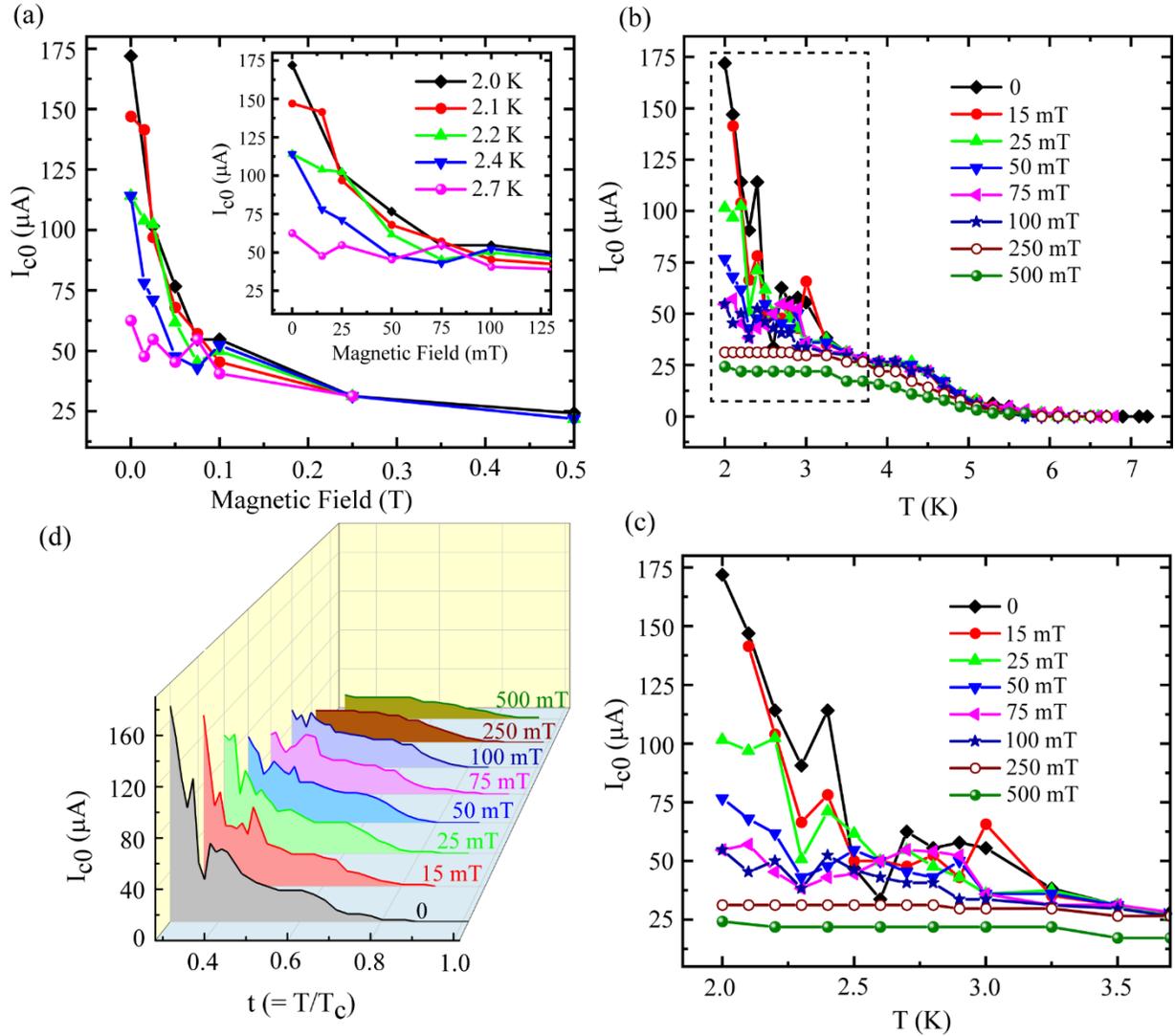

**Fig. 7:** Dependence of the critical current $I_{c0}$ on magnetic field and temperature. (a) Variation of $I_{c0}$ with magnetic field for specific temperatures. Inset: magnified view of the low-field region of the main panel. (b) $I_{c0}$ as a function of temperature for different magnetic field and the low temperature bounded region is zoomed in (c). (d) 3-D representation of $I_{c0}$ as a function of reduced temperature for all the magnetic fields shown in (b) & (c).

**Supplementary Material**

# Accessing Phase Slip Events in Nb Meander Wires

*Deepika Sawle[1,2], Sudhir Husale[1,3], Sachin Yadav[1,2], Bikash Gajar[1,2], V. P. S. Awana[1,2] and Sangeeta Sahoo[1,2]\**

[1]*Academy of Scientific and Innovative Research (AcSIR), AcSIR Headquarters CSIR-HRDC Campus, Ghaziabad, Uttar Pradesh, 201002, India.*

[2]*Electrical & Electronics Metrology Division, National Physical Laboratory, Council of Scientific and Industrial Research, Dr. K. S Krishnan Road, New Delhi-110012, India.*

[3]*Indian Standard Time Metrology Division, National Physical Laboratory, Council of Scientific and Industrial Research, Dr. K. S Krishnan Road, New Delhi-110012, India.*

*\*Correspondences should be addressed to S. S. (Email: sahoos@nplindia.org)*

## 1. The transport characteristics of the second meander sample

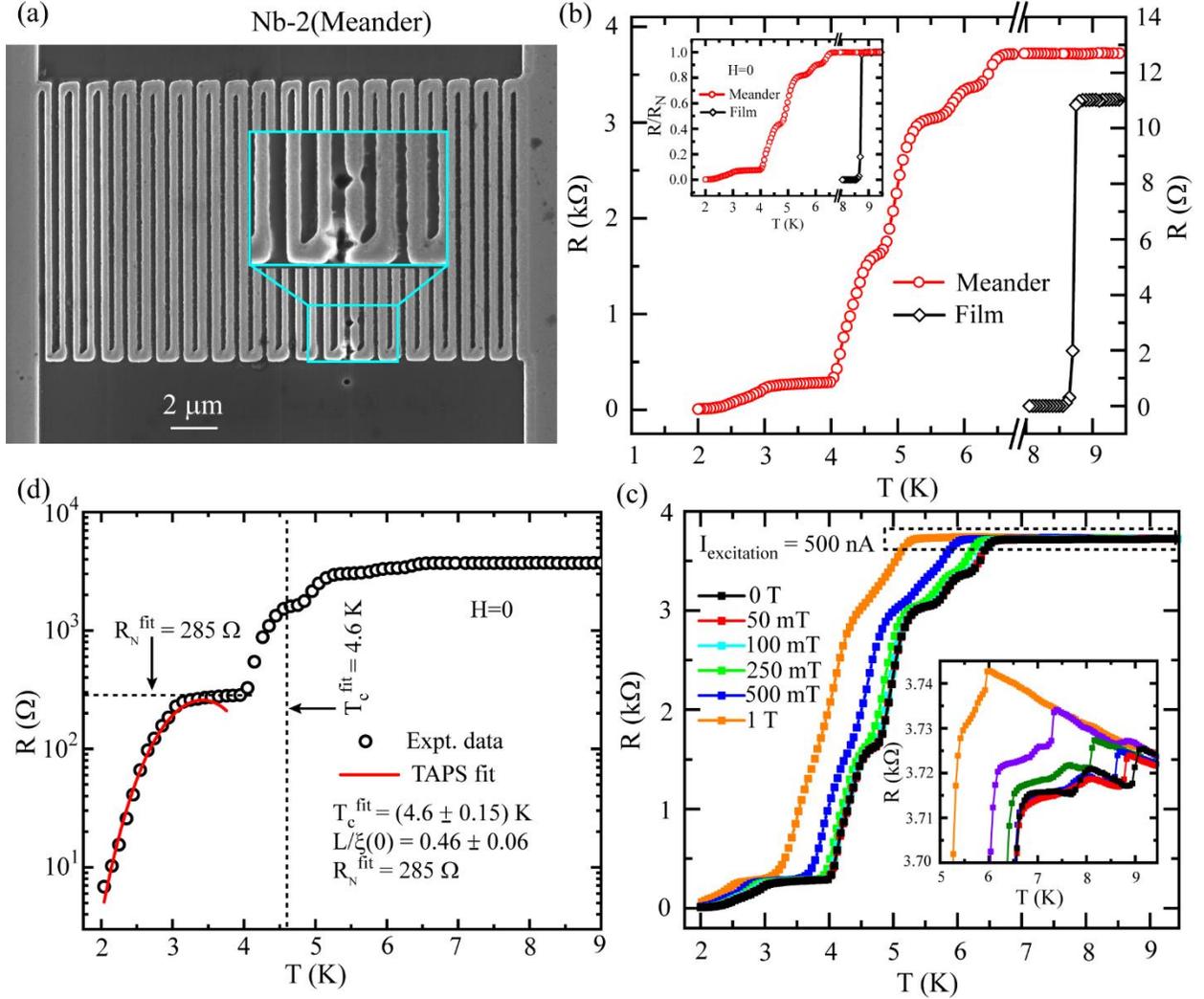

**Fig. S1:** Transport characteristics for the second meander sample Nb-2. (a) Scanning electron microscopy (SEM) image of the meander structure which reveals a region containing a couple of constrictions that resulted during the FIB milling. The damaged constriction containing region is highlighted at the center. (b) $R(T)$ comparison of thin film and meander by zero-field $R(T)$ measurements for a selected range of temperature before and after the meander fabrication. Inset: comparison of the normalized resistance for the film only and with the meander. Here, the resistance is normalized with respect to the normal state resistance at 10 K for both the configurations. (c) $R(T)$ measurements under magnetic field applied perpendicular to the sample plane. The region bounded by the dotted rectangular box is magnified in the inset. Here, a slight upturn in $R(T)$ before transition is evident. (d) A semi-log representation of zero-field $R(T)$ measured on the meander device along with a fit (the red curve) based on the thermally activated phase slip (TAPS) model using Eqn. 2 in the main manuscript.

Scanning electron microscopy (SEM) image for the second meander device is shown in Fig. S1(a), which reveals that a region of the meander was damaged during the focused ion beam (FIB) milling. The damaged portions look like constrictions and for better view high resolution image of the constrictions are placed on top of the meander in Fig. S1(a). Low temperature transport measurements were carried out on this device too before and after the fabrication of the meander structure. The temperature dependent resistance $R(T)$ measurements were performed down to 2 K and the comparison in $R(T)$ with respect to meandering is presented in Fig. S1(b). It clearly shows that normal state resistance of Nb meander is more than two orders of magnitude more than that of the Nb film. Moreover, single step sharp transition, occurred for the Nb thin film, gets modified into broad and multi-step transition from normal state to superconducting state for Nb meander. For the better comparison, both the $R(T)s$ were normalized by the respective resistance at 10 K as shown in the inset of Fig. S1(b). It clearly demonstrates the existence of resistive tailing occurring in the $R(T)$ of Nb meander at temperature low temperature ($< 4$ K). Further $R(T)$ measurements were also carried out in the presence of applied magnetic field as presented in Fig. S1(c). Here, multiple steps become smooth and transform into a continuously varying $R(T)$ characteristic with increasing magnetic field, whereas resistive tailing remains prominent as depicted in Fig. S1(c). In order to have closure look on the normal state behaviour under applied magnetic field, we have zoomed the selected region as bounded by the dotted rectangle and presented in the inset of Fig. S1(c). A weak upturn appears in the R(T) before the transition and the upturn gets facilitated by the application of magnetic field. Finally, the resistive tailing part from the zero-field $R(T)$ was studied through TAPS fitting by using

Eqn.2 based on Lamber-Ambegaokar-McCumber-Halperin (LAMH) model as explained in the main manuscript. As it is clear in Fig. S1(d) that the fit follows the experimental data very well and the corresponding $T_c^{fit}$ and the $\frac{L}{\xi(0)}$ values are mentioned in the figure. The effective length taking part into the phase slip process is about half the coherence length indicating that the phase fluctuation is dominant at the constrictions.

2. **Calculation of Ginzburg- Landau (GL) coherence length ($\xi_{GL}$) for Nb thin film and meander samples.**

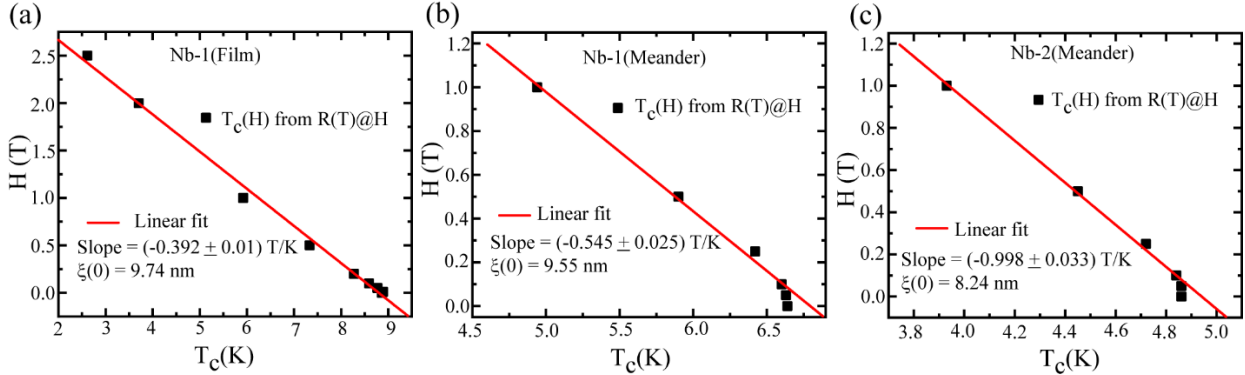

*Fig. S2: B-T phase diagram obtained from R(T) measurements carried out under external magnetic field for (a) the first sample (Nb-1) before meandering, (b) the same sample (Nb-1) after meandering and (c) the second sample (Nb-2) after meandering. Black scattering squares are the data points collected from temperature dependent resistance measurements under applied field [R(T)@H]. Solid red lines represent the linear fits performed on the experimental data points and provide the slope for calculating the GL coherence length $\xi_{GL}$.*

Here, we have calculated the Ginzburg-Landau (GL) coherence length $\xi_{GL}(0)$, by using the standard formula, $\xi_{GL}(0) = \left[\frac{\phi_0}{2\pi T_c \left|\frac{dH_{c2}}{dT}\right|_{T_c}}\right]^{1/2}$, where $\phi_0$ is the flux quantum. The experimental data points for the samples Nb-1(film) before meandering and Nb-1(Meander), Nb-2(Meander) after meandering are obtained from the $T_c$ values taken from the field dependent $R(T)$. The extracted values from the field dependent $R(T)$ are fitted linearly in Fig. S2 as shown by the red lines. The slopes obtained from the linear fits were used for calculating the coherence length $\xi_{GL}(0)$. Corresponding coherence lengths are 9.74 nm [Nb-1(film)], 9.55 nm [Nb-1(meander)] & 8.24 nm [Nb-2(meander)], respectively.